\documentclass[11pt,english]{article}
\usepackage[T1]{fontenc}
\usepackage[latin9]{inputenc}
\usepackage{geometry}
\usepackage{array}
\usepackage{longtable}
\usepackage{float}
\usepackage{textcomp}
\usepackage{multirow}
\usepackage{amsmath}
\usepackage{graphicx}
\usepackage{epstopdf}
\usepackage{setspace}
\usepackage{esint}

\makeatletter

\providecommand{\tabularnewline}{\\}

\newcommand{\lyxaddress}[1]{
\par {\raggedright #1
\vspace{1.4em}
\noindent\par}
}

\makeatother

\usepackage{babel}
\begin{document}

\title{Charged Scalar Perturbations around Garfinkle-Horowitz-Strominger Black
Holes}

\author{Cheng-Yong Zhang$^1$%
\thanks{zhangcy@sjtu.edu.cn%
}, Shao-Jun Zhang$^2$%
\thanks{sjzhang84@hotmail.edu.cn%
}, and Bin Wang$^1$%
\thanks{wang\_b@sjtu.edu.cn%
}}
\date{}
\maketitle
\lyxaddress{\begin{center} $^1$ Department of
Physics and Astronomy, Shanghai Jiao Tong
University, Shanghai 200240, China\\
$^2$ Instituto de
F$\acute{i}$sica, Universidade de S$\tilde{a}$o Paulo, C.P. 66318,
05315-970, S$\tilde{a}$o Paulo, SP, Brazil
\par\end{center}}

\begin{abstract}
We examine the stability of the
Garfinkle-Horowitz-Strominger (GHS)  black hole
under charged scalar perturbations. We find that
different from the neutral scalar field
perturbations, only two numerical methods, such
as the continued fraction method and the
asymptotic iteration method, can keep high
efficiency and accuracy requirements in the
frequency domain computations. The comparisons of
the efficiency between these two methods have
also been done. Employing the appropriate
numerical method, we show that the GHS black hole
is always stable against charged scalar
perturbations. This is different from the result
obtained in the de Sitter and Anti-de Sitter
black holes. Furthermore we argue that in the GHS
black hole background there is no amplification
of the incident charged scalar wave to cause the
superradiance, so that the superradiant
instability cannot exist in this spacetime.
\end{abstract}


\section{Introduction}
Perturbation around black holes has been an
intriguing subject of discussions in the past
three decades. This is mainly because that the
study of black hole perturbations is a powerful
tool to disclose the stability of the black hole
spacetime. If the black hole is unstable against
small perturbations, it will inevitably disappear
or transform dynamically into another object.
Stability analysis of (3+1)-dimensional
asymptotically flat black holes, such as
Schwarzschild, Reissner-Nordstrom (RN) and Kerr
black holes, has been studied thoroughly against
different kinds of perturbations including
neutral scalar, electromagnetic and gravitational
perturbations and all these black holes were
found stable. Considering that our universe may
have a small positive cosmological constant, the
perturbation analysis has been extended to
(3+1)-dimensional Schwarzschild-de Sitter (dS),
RN-dS, and Kerr-dS black holes and evidences for
the stability of these more possible
astrophysical black holes have been reported.
Recently, motivated by the discovery of the
correspondence between physics in the anti-de
Sitter (AdS) spacetime and conformal field theory
(CFT) on its boundary (AdS/CFT), the
perturbations around four-dimensional AdS black
holes have been examined and (3+1)-dimensional
AdS black holes were found stable under neutral
scalar, electromagnetic and gravitational
perturbations. It was concluded that all of the
considered four-dimensional black holes tested
for stability are stable, except the string
theory generalization of Kerr-Newman black holes
whose stabilities have not been tested due to the
difficulty in decoupling the angular variables in
their perturbation equations. For a review on
this topic, see \cite{Konoplya:2011,Wang:2005}
for example and references therein.

Recently, the exploration of the black hole
spacetime stability has been extended to examine
the perturbation against the charged scalar
field. In the AdS spacetime, it was first
observed that the (3+1)-dimensional RN-AdS black
hole can be destroyed and the four-dimensional
AdS hole can become unstable due to the
condensation of the charged scalar hair onto the
black hole. The (3+1)-dimensional RN-AdS black
hole will finally be transformed into another new
hairy black hole under the small charged scalar
field perturbation
\cite{XH:2010,Abdalla:2010,YL:2011}, see review
for example \cite{3H}. It would be interesting to
ask whether the observed instability only happens
for the AdS black holes because of their special
spacetime properties, whether such dynamical
instability can also appear in other
four-dimensional black hole backgrounds. In
\cite{Zhu:2014}, a new instability in the
four-dimensional RN-dS black holes against
charged scalar perturbations was disclosed. This
result was later confirmed in
\cite{Konoplya:2014lha}. Can this instability be
a general property? In this paper, we would like
to examine this problem further. We will extend
the discussion of the charged scalar perturbation
to the (3+1)-dimensional dilaton black hole
obtained by Garfinkle, Horowitz and
Strominger~\cite{Garfinkle:1991ghs}(GHS) from the
low energy effective action in string theory. Its
limiting case reduces to the four-dimensional
Schwarzschild black hole, see review
~\cite{Horowitz:1992jp}. The stability of this
dilaton black hole has been proved by examining
the perturbation against neutral scalar fields
~\cite{dilaton:low,dilaton:high,dilaton:4}. Here
we are going to test the stability of such black
hole against charged scalar perturbation and try
to answer whether the instability observed for
the charged scalar perturbation can also happen
in this stringy four-dimensional black hole
background and its limiting Schwarzschild
spacetime.

We will concentrate on the frequency domain
studies of the charged scalar perturbation around
the stringy black hole. It is important to
calculate the frequency of perturbations with
very high accuracy because considerable changing
of black hole parameters frequently changes
perturbation frequency just by a few percent.
With the nonzero charge of the scalar field, we
will explain that not all available numerical
methods for solving the eigenvalue problem of the
perturbation can keep high accuracy. The
continued fraction method (CFM)
\cite{Leaver,Leaver:1990RN} and the asymptotic
iteration method (AIM)
\cite{Ciftci,Cho:2010,Cho:2012} still win the
accuracy and efficiency competition against the
other numerical methods.  The CFM method was
argued as the most accurate in calculating the
frequency of the perturbation
\cite{Cardoso:0905}. Although this still holds
when the charge of the scalar field is low, we
will show that its accuracy and efficiency will
be decreased and lower than that of the AIM
method when the scalar field is heavily charged.
Further comparison of the accuracy and efficiency
of numerical methods in calculating the frequency
of the perturbation is important, because this
can help to choose the right numerical tool in
obtaining the real physics on the stability of
the black hole.

The organization of the paper is as follows. In
Sec.2 we will introduce the background spacetime
of the stringy black hole and derive the equation
of motion for charged scalar perturbations. In
Sec.3, we will first review the CFM and AIM
methods for numerical computations of the
frequency of the perturbation. Then we will
compare the efficiency and accuracy of these two
methods and also with other methods. In the
following section, we we will give numerical
results on the frequency of the charged scalar
perturbations. We will present our summaries and
discussions in the final section.

\section{The GHS black hole and equation of charged scalar perturbation}

The GHS black hole is a solution obtained from
the low energy effective action in string theory
by dropping all the fields except the metric
$g_{\mu\nu}$, a dilaton $\phi$ and a Maxwell
field $F_{\mu\nu}$. In string frame, the action
is~\cite{Frolov:1998}
\begin{equation}
W=\frac{1}{16\pi}\int d^{4}x\sqrt{-g}e^{-2\phi}[R+4(\nabla\phi)^{2}-F_{\mu\nu}F^{\mu\nu}]-\frac{1}{2}g^{\mu\nu}\bar{D}_{\mu}\bar{\psi}D_{\nu}\psi-V(\psi,\bar{\psi}),
\end{equation}
where we have added the perturbing charged scalar
field $\psi$ to study its perturbation on the
background of the GHS black hole. Here
$\bar{D}_{\mu}\bar{\psi}D_{\nu}\psi \equiv
(\partial_{\mu}+iqA_{\mu})\bar{\psi}(\partial_{\nu}-iqA_{\nu})\psi$
with $q$ being the charge of the scalar field
$\psi$. $V(\psi,\bar{\psi})$ is the potential of
the perturbing charged scalar field. Its usual
form is taken as
$V(\psi,\bar{\psi})=\frac{1}{2}\mu^{2}\psi\bar{\psi}+\frac{\lambda}{4}(\psi\bar{\psi})^{2}$,
in which $\mu$ is the mass of the scalar field
$\psi$ and the $\lambda$ term represents the
self-interaction of $\psi$. We set $\lambda=0$
hereafter for simplicity. This action can be
viewed as a special case in scalar-tensor
theory~\cite{Fujii:2003} with
$F(\phi)=e^{-2\phi}$ and $Z(\phi)=-4e^{-2\phi}$.

Doing a conformal transformation
$g_{\mu\nu}^{E}=e^{-2\phi}g_{\mu\nu}$, the action
can be rewritten in the Einstein frame as
\begin{equation}
W=\frac{1}{16\pi}\int d^{4}x\sqrt{-g_{E}}[R_{E}-2(\nabla\phi)^{2}-e^{-2\phi}F_{\mu\nu}F^{\mu\nu}]-\frac{1}{2}e^{2\phi}g_{E}^{\mu\nu}\bar{D}_{\mu}\bar{\psi}D_{\nu}\psi-e^{4\phi}V(\psi,\bar{\psi}).\label{eq:action-e}
\end{equation}
At the first sight, it seems difficult to find
the exact background solution to equations of
motion derived from the action
Eq.~(\ref{eq:action-e}) (The background solution
means the solution not taking into account the
back-reaction of the perturbing field $\psi$).
However, the symmetry property of this action
allows one to obtain a one-parameter family of
solutions~\cite{Garfinkle:1991ghs,Horowitz:1992jp,Gibbons:1982},
\begin{equation}
ds_{E}^{2}=-\left(1-\frac{2M}{r}\right)dt^{2}+\left(1-\frac{2M}{r}\right)^{-1}dr^{2}+r\left(r-\frac{Q^{2}}{M}\right)d\Omega^{2}.
\end{equation}
This is the well-known
Garfinkle-Horowitz-Strominger (GHS) black hole.
Here $d\Omega^{2}=d\theta^{2}+\sin^{2}\theta
d\varphi^{2}$. $M$ is the physical mass and $Q$
the physical charge of the GHS black hole. The
electric field and the background dilaton field
are
\begin{eqnarray}
A_{t}=-\frac{Q}{r}, & F_{rt}=\frac{Q}{r^{2}}, & e^{2\phi}=1-\frac{Q^{2}}{Mr}.
\end{eqnarray}
It is obvious that the electric charge and the
dilaton are not independent. The GHS solution
reduces to the Schwarzschild black hole in the
limit $Q\rightarrow 0$.

In the limit $Q\rightarrow 0$, the area of the
sphere $r=Q^{2}/M$ is zero so that this surface
is singular. When $Q^{2}<2M^{2}$, this singular
surface is surrounded by the event horizon
$r=2M$. As one increases $Q$, the singular
surface can coincide with the horizon when
$Q^{2}=2M^{2}$ and even moves outside the horizon
and becomes timelike if $Q^{2}>2M^{2}$. In our
paper, we will consider the case when
$Q^{2}<2M^{2}$.

Ignoring the self-interaction term, the equation of motion of the perturbing
charged scalar field is
\begin{equation}
\left(g_{E}^{\mu\nu}D_{\mu}D_{\nu}-\mu^{2}e^{2\phi}\right)\psi=0.
\end{equation}
Here $\mu^{2}e^{2\phi}$ plays the role of the
effective mass square of $\psi$. Since the
background spacetime is spherical symmetry, we
can separate the radial and angular part of
$\psi$. Using the ansatz $\psi=e^{-i\omega
t}\frac{\Psi(r)}{\sqrt{r^{2}-\frac{Q^{2}}{M}r}}Y(\theta,\varphi)$
and introducing the tortoise coordinate
$dr=\left(1-\frac{2M}{r}\right)dr_{*}$, we get
the radial part of the perturbation equation
\begin{eqnarray}
\frac{\partial^{2}\Psi}{\partial r_{*}^{2}}+\left[\left(\omega-\frac{qQ}{r}\right)^{2}-V\right]\Psi = 0 \label{eq:radial_eq}
\end{eqnarray}
in which
\begin{eqnarray}
V & = & -\frac{3}{4}\frac{\left(2r-\frac{Q^{2}}{M}\right)^{2}\left(r-2M\right)^{2}}{r^{4}\left(r-\frac{Q^{2}}{M}\right)^{2}}+\frac{\left(r-2M\right)^{2}}{r^{3}\left(r-\frac{Q^{2}}{M}\right)}+\frac{\left(2r-\frac{Q^{2}}{M}-2M\right)\left(2r-\frac{Q^{2}}{M}\right)\left(r-2M\right)}{2\left(r-\frac{Q^{2}}{M}\right)^{2}r^{3}}\nonumber \\
 &  & +\frac{\left(r-\frac{Q^{2}}{M}\right)\left(r-2M\right)}{r^{2}}\mu^{2}+\frac{l(l+1)\left(r-2M\right)}{\left(r-\frac{Q^{2}}{M}\right)r^{2}}
\end{eqnarray}
and the effective potential $V_{eff}=V-\frac{q^{2}Q^{2}}{r^2}$. Here $l$ is the spherical harmonic index. It can be shown that we
always have $V>0$ when $r>2M$ and $Q^2 <2 M^2$.

\section{Numerical methods}

A practical tool for testing stability of black
holes is the numerical investigation of the
perturbations around black hole backgrounds.
Usually considerable changing of black hole
parameters results in the change of just a few
percent in the frequency of the perturbation.
Thus the high accuracy of the computation is the
key factor in examining the perturbation around
black holes. Meanwhile the efficiency of the
computation is also an important factor in
solving the perturbation equations numerically.

In this work, we will concentrate on the
frequency domain to disclose the property of the
perturbation. We will refine different numerical
methods and try to solve the eigenvalue problem
of the perturbation equation with high accuracy
and efficiency. Comparing with other numerical
methods, we find that the CFM and AIM methods can
meet requirements of the high accuracy and
efficiency in the computation. We will first
review these two methods. Furthermore we will
show that in different parameter ranges, the
accuracy and efficiency also differ between these
two refined methods. Without loss of generality,
we will set $M=1$ and $\mu=0$ in the following
discussions.

\subsection{Continued fraction method}

The CFM  was proposed by Leaver~\cite{Leaver}
when he calculated the quasinormal modes (QNMs)
of the Kerr black hole and is considered as the
most accurate method to calculate the frequencies
of perturbations~\cite{Cardoso:0905}. The core of
this method is to cast the perturbation equation
into a three-term recurrence relation, and from
it we can get a continued fraction equation
characterizing the perturbations. The CFM is
thought to be able to give frequencies of
perturbations with high numerical precision as
there is no intermediate approximation compared
to other numerical methods. See
reviews~\cite{Konoplya:2011,Cardoso:0905} for more
details. In this subsection, we try to get the
three-term recurrence relation and the
corresponding continued fraction equation.

To calculate the frequency, we start from the
physical boundary conditions which can be derived
by studying the asymptotic behavior of
Eq.~(\ref{eq:radial_eq})
\begin{equation}
\Psi\sim\begin{cases}
e^{-i\left(\omega-\frac{qQ}{2M}\right)r_{*}}, & r_{*}\rightarrow-\infty(r\rightarrow2M),\\
e^{i\sqrt{\omega^{2}-\mu^{2}}r_{*}}, & r_{*}\rightarrow\infty(r\rightarrow\infty),
\end{cases}
\end{equation}
which means that there only exists the ingoing
wave at the event horizon and the outgoing wave
at the infinity.

A solution to the radial equation
Eq.~(\ref{eq:radial_eq}) encoding the above
boundary conditions can be written in the form as
\begin{equation}
\Psi=\left(\frac{r}{2M}-1\right)^{-i(2M\omega-qQ)}\left(\frac{r}{2M}\right)^{i2M(\omega+\sqrt{\omega^{2}-\mu^{2}}-\frac{qQ}{2M})}e^{i\sqrt{\omega^{2}-\mu^{2}}(r-2M)}\sum_{n=0}^{\infty}a_{n}x^{n},
\end{equation}
in which $x=\frac{r-2M}{r}$ and $a_{0}=1$. Substituting this expansion into the radial equation (\ref{eq:radial_eq}),
we get a six-term recurrence relation.
\begin{eqnarray}
\beta_{0}a_{0}+\alpha_{0}a_{1} & = & 0,\nonumber \\
\gamma_{1}a_{0}+\beta_{1}a_{1}+\alpha_{1}a_{2} & = & 0,\nonumber \\
\delta_{2}a_{0}+\gamma_{2}a_{1}+\beta_{2}a_{2}+\alpha_{2}a_{3} & = & 0,\\
\eta_{3}a_{0}+\delta_{3}a_{1}+\gamma_{3}a_{2}+\beta_{3}a_{3}+\alpha_{3}a_{4} & = & 0,\nonumber \\
\theta_{4}a_{0}+\eta_{4}a_{1}+\delta_{4}a_{2}+\gamma_{4}a_{3}+\beta_{4}a_{4}+\alpha_{4}a_{5} & = & 0,\nonumber \\
\theta_{n}a_{n-4}+\eta_{n}a_{n-3}+\delta_{n}a_{n-2}+\gamma_{n}a_{n-1}+\beta_{n}a_{n}+\alpha_{n}a_{n+1} & = & 0,\nonumber
\end{eqnarray}
where the recurrence coefficients are given by
 \begin{eqnarray}
\alpha_{n} & = & (Q^{2}-2)^{2}\left[-4n^{2}-8\left(1+iqQ-2i\omega\right)n-4(1+2iqQ)+16i\omega\right],\nonumber \\
\beta_{n} & = & 2(Q^{2}-2)\left[2n^{2}(5Q^{2}-6)+4n\left(-2+iqQ(5Q^{2}-6)+Q^{2}(1-14i\omega)+20i\omega\right)\right] \\
 &  & +2(Q^{2}-2)\left[-4-4l(l+1)+Q^{2}+4qQ(i-qQ)(Q^{2}-2)\right]\nonumber \\
 &  & +2(Q^{2}-2)\left[8(5qQ-2i)(Q^{2}-2)\omega-64(Q^{2}-2)\omega^{2}\right],\nonumber
\end{eqnarray}
\begin{eqnarray}
 \gamma_{n} & = & -8n^{2}(5Q^{4}-12Q^{2}+6)+16l(l+1)(Q^{2}-1)\nonumber \\
 &  & -16+36Q^{2}-17Q^{4}+16iqQ(2-6Q^{2}+3Q^{4})+4q^{2}Q^{2}(12-20Q^{2}+7Q^{4})\nonumber \\
 &  & -16\left[i(8-20Q^{2}+9Q^{4})+4qQ(6-11Q^{2}+4Q^{4})\right]\omega+64(12-20Q^{2}+7Q^{4})\omega^{2}\nonumber \\
 &  & -16in\left[2i+qQ(6-12Q^{2}+5Q^{4})+Q^{4}(3i-17\omega)-24\omega+Q^{2}(44\omega-6i)\right],\nonumber\\
\delta_{n} & = & 8n^{2}(2-8Q^{2}+5Q^{4})-8l(l+1)Q^{2}\label{eq:cofficient}\\
 &  & +16-100Q^{2}+79Q^{4}-16iqQ(2-10Q^{2}+7Q^{4})-4q^{2}Q^{2}(4-16Q^{2}+9Q^{4})\nonumber \\
 &  & +16\left[2i(4-20Q^{2}+13Q^{4})+qQ(8-32Q^{2}+19Q^{4})\right]\omega-64(4-16Q^{2}+9Q^{4})\omega^{2}\nonumber \\
 &  & +16in\left[2i+qQ(2-8Q^{2}+5Q^{4})+Q^{4}(7i-19\omega)-8\omega+2Q^{2}(16\omega-5i)\right],\nonumber
\end{eqnarray}
 \begin{eqnarray}
\eta_{n} & = & Q^{2}\left[-4n^{2}(5Q^{2}-4)+4q^{2}Q^{2}(5Q^{2}-4)\right]\nonumber \\
 &  & +8nQ^{2}\left[-8-iqQ(5Q^{2}-4)+Q^{2}(11+20i\omega)-16i\omega\right] \\
 &  & +Q^{2}(-5i+8\omega)\left[12i-32\omega+Q^{2}(-19i+40\omega)\right]\nonumber \\
 &  & +8qQ^{3}\left[8i-16\omega+Q^{2}(-11i+20\omega)\right],\nonumber \\
\theta_{n} & = & Q^{4}\left[4n^{2}+8in(3i+qQ-4\omega)+35-4q^{2}Q^{2}+96i\omega-64\omega^{2}+8qQ(-3i+4\omega)\right].\nonumber
\end{eqnarray}

% Here we have made $M=1$ and $\mu=0$ for simplicity.
We can use the Gauss elimination to reduce the six-term recurrence relation
to a five-term recurrence relation,
\begin{eqnarray}
\beta'_{0}=\beta_{0},\gamma'_{1}=\gamma{}_{1},\beta'_{1}=\beta_{1} & , & \delta'_{2}=\delta_{2},\gamma'_{2}=\gamma_{2},\beta'_{2}=\beta_{2},\nonumber \\
\eta'_{3}=\eta_{3},\delta'_{3}=\delta_{3},\gamma'_{3}=\gamma_{3} & , & \beta'_{3}=\beta_{3},\alpha'_{n}=\alpha{}_{n}(n\geq0),\\
\eta'_{n}=\eta_{n}-\frac{\theta_{n}}{\eta'_{n-1}}\delta'_{n-1} & , & \delta'_{n}=\delta_{n}-\frac{\theta_{n}}{\eta'_{n-1}}\gamma'_{n-1}(n\geq4),\nonumber \\
\gamma'_{n}=\gamma_{n}-\frac{\theta_{n}}{\eta'_{n-1}}\beta'_{n-1} & , & \beta'_{n}=\beta_{n}-\frac{\theta_{n}}{\eta'_{n-1}}\alpha'_{n-1}(n\geq4).\nonumber
\end{eqnarray}
By repeating the Gauss elimination, a four-term recurrence relation can be
derived,
\begin{eqnarray}
\beta''_{0}=\beta'_{0},\gamma''_{1}=\gamma'_{1} & , & \beta''_{1}=\beta'_{1},\nonumber \\
\delta''_{2}=\delta'_{2},\gamma''_{2}=\gamma'_{2},\beta''_{2}=\beta'_{2} & , & \alpha''_{n}=\alpha'{}_{n}(n\geq0),\\
\delta''_{n}=\delta'_{n}-\frac{\eta'_{n}}{\delta''_{n-1}}\gamma''_{n-1},\gamma''_{n}=\gamma'_{n}-\frac{\eta'_{n}}{\delta''_{n-1}}\beta''_{n-1} & , & \beta''_{n}=\beta'_{n}-\frac{\eta'_{n}}{\delta''_{n-1}}\alpha''_{n-1}(n\geq3).\nonumber
\end{eqnarray}
And at last we get a three-term recurrence relation
\begin{eqnarray}
\beta'''_{0}=\beta''_{0},\gamma'''_{1}=\gamma''_{1} & , & \beta'''_{1}=\beta''_{1},\alpha'''_{n}=\alpha''{}_{n}(n\geq0),\\
\gamma'''_{n}=\gamma''_{n}-\frac{\delta''_{n}}{\gamma'''_{n-1}}\beta'''_{n-1} & , & \beta'''_{n}=\beta''_{n}-\frac{\delta''_{n}}{\gamma'''_{n-1}}\alpha'''_{n-1}(n\geq2).\nonumber
\end{eqnarray}
Then the frequencies of the perturbations are the
solutions to the characteristic continued
fraction equation
\begin{equation}
0=\beta'''_{0}-\frac{\alpha'''_{0}\gamma'''_{1}}{\beta'''_{1}-}\frac{\alpha'''_{1}\gamma'''_{2}}{\beta'''_{2}-}\frac{\alpha'''_{2}\gamma'''_{3}}{\beta'''_{3}-\cdots}.
\end{equation}
It is obvious that the six-term recurrence
relation reduces to a four-term recurrence
relation when $Q=0$ or $Q=\sqrt{2}$ due to the
vanish of some coefficients in
(\ref{eq:cofficient}). When $Q=0$, the GHS metric
reduces to the Schwarzschild metric, and the
six-term recurrence relation boils down to a
four-term recurrence relation which later
coincides with the three-term recurrence relation
derived in ref.~\cite{Leaver} after doing the
Gauss elimination with $2M=1$.

\subsection{Asymptotic iteration method}

The asymptotic iteration method (AIM) was first
used to solve the eigenvalue problems of second
order homogeneous linear differential
equations~\cite{Ciftci}. It was then applied to
find the frequencies of perturbations in
Schwarzschild and Schwarzschild (anti-) de Sitter
black holes~\cite{Cho:2010}. See
review~\cite{Cho:2012} and references therein.

Let's consider a second order homogeneous linear
differential equations
\begin{equation}\label{standardform}
\chi''=\lambda_{0}(x)\chi'+s_{0}(x)\chi,
\end{equation}
where $\lambda_{0}(x)$ and $s_{0}(x)$ are smooth functions in some
interval $[a,b]$. Differentiating it with respect to $x$, we get
\begin{equation}
\chi'''=\lambda_{1}(x)\chi'+s_{1}(x)\chi,
\end{equation}
 where
\begin{equation}
\lambda_{1} (x) =\lambda_{0}'(x)+s_{0}(x)+(\lambda_{0}(x))^{2}, \quad s_{1}=s_{0}'(x)+s_{0}(x)\lambda_{0}(x).
\end{equation}
 Using this step iteratively, we can get the $(n+2)$th derivatives
\begin{equation}
\chi^{(n+2)}=\lambda_{n}(x)\chi'+s_{n}(x)\chi
\end{equation}
 where
\begin{equation}
\lambda_{n}(x)=\lambda'_{n-1}(x)+s_{n-1}(x)+\lambda_{0}(x)\lambda_{n-1}(x),s_{n}(x)=s'_{n-1}(x)+s_{0}(x)\lambda_{n-1}(x).\label{eq:AIM_iteration}
\end{equation}
 For sufficiently large $n$, the asymptotic aspect of the method
was introduced~\cite{Cho:2012},
\begin{equation}
\frac{s_{n}(x)}{\lambda_{n}(x)}=\frac{s_{n-1}(x)}{\lambda_{n-1}(x)}
\end{equation}
 which is equivalent to imposing a termination to the number of iterations~\cite{Barakat:2006aim}.
The perturbation frequencies can be derived from
this ``quantization condition''. However, the
derivatives of $\lambda_{n}(x)$ and $s_{n}(x)$ in
each iteration slow down the AIM considerably and
also lead to precision problems. These drawbacks
were overcomed in ref.~\cite{Cho:2010}. One can
expand $\lambda_{n}(x)$ and $s_{n}(x)$ in Taylor
series around a regular point $\xi$ at which the
AIM is performed,
\begin{eqnarray}
\lambda_{n}(\xi)=\sum_{i=0}^{\infty}c_{n}^{i}(x-\xi)^{i} & , & s_{n}(x)=\sum_{i=0}^{\infty}d_{n}^{i}(x-\xi)^{i}.\label{eq:AIM_expansion}
\end{eqnarray}
 Here $c_{n}^{i}$ and $d_{n}^{i}$ are the $i$th Taylor coefficients
of $\lambda_{n}(\xi)$ and $s_{n}(\xi)$, respectively. Substituting
these expansions into (\ref{eq:AIM_iteration}), we get a set of recursion
relations for the coefficients,
\begin{eqnarray}
c_{n}^{i}=(i+1)c_{n-1}^{i+1}+d_{n-1}^{i}+\sum_{k=0}^{i}c_{0}^{k}c_{n-1}^{i-k} & , & d_{n}^{i}=(i+1)d_{n-1}^{i+1}+\sum_{k=0}^{i}d_{0}^{k}c_{n-1}^{i-k}.\label{eq:AIM_exp_iteration}
\end{eqnarray}
 The quantization condition then can be expressed as
\begin{equation}
d_{n}^{0}c_{n-1}^{0}-d_{n-1}^{0}c_{n}^{0}=0,\label{eq:quantization}
\end{equation}
 which will give us the perturbation frequencies of a black hole. Both the accuracy and
efficiency of AIM are greatly improved with this
expansion~\cite{Cho:2010}.

Now we apply this method to the GHS black hole.
Taking a coordinate transformation
$x=1-\frac{2M}{r}$ and an abbreviation
$a=\frac{Q^{2}}{2M^{2}}$, the perturbation
equation (\ref{eq:radial_eq}) turns into the
standard form as Eq.~(\ref{standardform})
\begin{equation}
\frac{\partial^{2}\Psi}{\partial x^{2}}=\lambda_{0}(x)\frac{\partial\Psi}{\partial x}+s_{0}(x)\Psi
\end{equation}
in which
\begin{eqnarray}
\lambda_{0}(x) & = & \frac{3x-1}{x(1-x)},\label{eq:GHS_AIM_cof}\\
s_{0}(x) & = & \frac{-1}{x^{2}(1-x)^{2}}\left[\frac{3}{4}\left(\frac{2-a+ax}{1-a+ax}\right)^{2}x^{2}-\frac{x^{2}}{1-a+ax}-\frac{1}{2}\left(\frac{1-a+(a+1)x}{1-a+ax}\right)\left(\frac{2-a+ax}{1-a+ax}\right)x\right]\nonumber \\
 &  & -\frac{1}{x^{2}(1-x)^{2}}\left[\left(\frac{2M\omega}{1-x}-qQ\right)^{2}-\left(1-a+ax\right)\frac{4M^{2}x}{(1-x)^{2}}\mu^{2}-\frac{l(l+1)x}{\left(1-a+ax\right)}\right].\nonumber
\end{eqnarray}
Now the infinity corresponds to $x\rightarrow1$
and the horizon is at $x\rightarrow0$. We can
choose the regular point $\xi$ between $0$ and
$1$. Substituting~(\ref{eq:GHS_AIM_cof}) into the
(\ref{eq:AIM_expansion})
(\ref{eq:AIM_exp_iteration})
(\ref{eq:quantization}), we can get  frequencies
of perturbations around the GHS black hole.

As we can see from (\ref{eq:AIM_expansion}), the
efficiency and accuracy of the numerical result
depend on the position of the expansion $\xi$. In
our calculation, we find that when the charge of
the black hole is not too large, the position of
the expansion point has little influence and the
AIM converges well. However, when the charge of
the black hole is larger than $M$, the position
of expansion will affect the efficiency and
accuracy of the numerical computation apparently.

\subsection{The efficiency and accuracy comparisons between AIM and CFM}

In this subsection, we present the efficiency and
accuracy comparisons between two numerical
methods, the CFM and the AIM. The reason for us
to concentrate on these two numerical methods in
doing computation is that we have found that
other numerical methods, such as the shooting
method~\cite{Chandrasekhar:1975,Molina:2010}, the
WKB
method~\cite{Schutz:1985,Iyer:1987,Konoplya:2004}
and the finite difference
method~\cite{Gundlach:1994,Wang:2004,Berti:2007}
cannot give us good convergence and reliability
in the computation when the scalar field is
charged, although they can give consistent
frequencies for the neutral scalar perturbations.
In our numerical computations we have set $M=1$
and $\mu=0$.

For the fundamental modes of the charged scalar
perturbation, it is  easy to  see that  both
methods, the AIM and the CFM, are easy to
converge. For example as shown in Table 1, when
$l=1,Q=0.5$ and $q=1$, only 20 iterative steps
are needed for CFM to get the fundamental mode
with relative error $10^{-5}$ compared to the
result obtained using 100 iterative steps, and
$30$ iterative steps for AIM with relative error
$10^{-4}$ compared to the results obtained by 100
iterative steps, respectively. However, for the
overtones, we found that more iterative steps are
needed compared to the fundamental modes in order
to keep the accuracy. We adopt 100 iterative
steps for both AIM and CFM in our numerical
calculations.

\begin{table}[h]
\centering
{\footnotesize{}}%
\begin{tabular}{|>{\centering}m{6.5cm}|>{\centering}m{3.5cm}|>{\centering}m{3.5cm}|}
\hline
{\footnotesize{}$l=1,Q=0.5,q=1$} & {\footnotesize{}CFM} & {\footnotesize{}AIM ($\xi=0.45$)}\tabularnewline
\hline
{\footnotesize{}iterative steps } & {\footnotesize{}20} & {\footnotesize{}30}\tabularnewline
\hline
{\footnotesize{}fundamental $n=0$}{\footnotesize \par}

{\footnotesize{}(relative error with respect to 100 steps)} & {\footnotesize{}0.491714-0.111583i}{\footnotesize \par}

{\footnotesize{}(0.0005\%,0.0028\%)} & {\footnotesize{}0.491758-0.111572i}{\footnotesize \par}

{\footnotesize{}(-0.0085\%,0.0128\%)}\tabularnewline
\hline
{\footnotesize{}overtone $n=1$}{\footnotesize \par}

{\footnotesize{}(relative error with respect to 100 steps)} & {\footnotesize{}0.479728-0.339701i}{\footnotesize \par}

{\footnotesize{}(0.0194\%,-0.0126\%)} & {\footnotesize{}0.477759-0.339279}{\footnotesize \par}

{\footnotesize{}(4.2983\%,0.1115\%)}\tabularnewline
\hline
\end{tabular}

{\footnotesize{}\protect\caption{\label{tab:
steps}{\footnotesize{} Fundamental mode $n=0$ and
the first overtone $n=1$ when $l=1, Q=0.5$ and
$q=1$. The two numbers in brackets are relative
errors of the real and imaginary part of the
modes compared to the ones calculated with 100
iteration steps, respectively. For the AIM, we do
the expansion at the point $\xi=0.45$.}} }
\end{table}

When the charge $Q$ of the black hole is small,
results given by two methods, the CFM and AIM,
agree quite well with each other. However, the
situation changes when $Q$ becomes large. We find
that for the chosen large value of $Q$, the speed
of convergence of AIM is faster than that of the
CFM when the scalar perturbation field is weakly
charged, but when the charge $q$ of the scalar
field increases,  the speed of the convergence of
the CFM becomes faster than that of the AIM. See
Table 2 for concrete examples.

\begin{table}[H]
\centering
{\footnotesize{}}%
\begin{tabular}{|>{\centering}m{1.2cm}|>{\centering}m{1.8cm}|>{\centering}m{6.3cm}|>{\centering}m{6.3cm}|}
\hline
\multicolumn{2}{|c|}{{\footnotesize{}$l=1,n=1$}} & \multicolumn{1}{c|}{{\footnotesize{}$q=0.5$}} & \multicolumn{1}{c|}{{\footnotesize{}$q=2$}}\tabularnewline
\hline
\multicolumn{2}{|c|}{{\footnotesize{}iterative steps }} & {\footnotesize{}80}{\footnotesize \par}

{\footnotesize{}(relative error with respect to 100 steps)} & {\footnotesize{}80}{\footnotesize \par}

{\footnotesize{}(relative error with respect to 100 steps)}\tabularnewline
\hline
\multirow{2}{1.2cm}{{\footnotesize{}$Q=0.1$}} & {\footnotesize{}AIM}{\footnotesize \par}

{\footnotesize{}($\xi=0.45$)} & {\footnotesize{}0.28448-0.31036i}{\footnotesize \par}

{\footnotesize{}($-4.1\times10^{-6},-3.0\times10^{-6}$)} & {\footnotesize{}0.343659-0.321025i}{\footnotesize \par}

{\footnotesize{}($-2.7\times10^{-6},-1.4\times10^{-7}$)}\tabularnewline
\cline{2-4}
 & {\footnotesize{}CFM} & {\footnotesize{}0.28448-0.31036i}{\footnotesize \par}

{\footnotesize{}($-2.1\times10^{-7},4.0\times10^{-6}$)} & {\footnotesize{}0.343658-0.321024i}{\footnotesize \par}

{\footnotesize{}($1.3\times10^{-6},3.4\times10^{-6}$)}\tabularnewline
\hline
\multirow{2}{1.2cm}{{\footnotesize{}$Q=1$}} & {\footnotesize{}AIM}{\footnotesize \par}

{\footnotesize{}($\xi=0.45$)} & {\footnotesize{}0.54590-0.34803i}{\footnotesize \par}

{\footnotesize{}($6.2\times10^{-8},-4.1\times10^{-7}$)} & {\footnotesize{}1.198949-0.380392i}{\footnotesize \par}

{\footnotesize{}($7.5\times10^{-6},1.6\times10^{-5}$)}\tabularnewline
\cline{2-4}
 & {\footnotesize{}CFM} & {\footnotesize{}0.54590-0.34803i}{\footnotesize \par}

{\footnotesize{}($-1.4\times10^{-6},3.2\times10^{-7}$)} & {\footnotesize{}1.198958-0.380398i}{\footnotesize \par}

{\footnotesize{}($-1.7\times10^{-11},3.6\times10^{-11}$)}\tabularnewline
\hline
\end{tabular}

{\footnotesize{}\protect\caption{\label{tab:Convergence}
{\footnotesize{}Convergence of CFM and AIM as
parameters $Q$ and $q$ vary. When $Q$ is small,
$Q=0.1$ for example, results given by both of
these two methods have high accuracy and agree
quite well. However, when $Q$ becomes large,
$Q=1$ for example, the convergence of the two
methods depends on the value of $q$. For small
$q$, $q=0.5$ for example, the speed of the
convergence of the AIM is faster than that of the
CFM. However, as $q$ increases, the speed of the
convergence of the AIM slows down, while the
speed of the convergence of the CFM increases and
becomes faster than that of the AIM.}} }
\end{table}

We have made comparisons of these two methods in
various situations, and the results are
summarized in Table 3. The AIM and the CFM have
little difference in the speed of convergence and
accuracy for small $Q$. For large $Q$ (but not
larger than $1$), the AIM converges faster than
the CFM when $q$ is small but when $q$ is large
we find that the CFM is a better way in
computation. When $Q$ is larger than $1$, the CFM
fails to converge and we can only rely on the
AIM.

\begin{table}[H]
\centering
\begin{tabular}{|>{\raggedright}m{2.2cm}|>{\centering}m{2.7cm}|>{\centering}p{2.7cm}|>{\centering}p{2.7cm}|>{\centering}p{2.7cm}|}
\hline
\multirow{2}{2.2cm}{{\footnotesize{}The speed of convergence}} & \multicolumn{2}{c|}{{\footnotesize{}Small $Q$}} & \multicolumn{2}{c|}{{\footnotesize{}large $Q$}}\tabularnewline
\cline{2-5}
 & {\footnotesize{}AIM} & {\footnotesize{}CFM} & {\footnotesize{}AIM} & {\footnotesize{}CFM}\tabularnewline
\hline
{\footnotesize{}Small $q$} & \multicolumn{2}{c|}{{\footnotesize{}Little difference}} & {\footnotesize{}Faster} & {\footnotesize{}Slower}\tabularnewline
\hline
{\footnotesize{}Large $q$} & \multicolumn{2}{c|}{{\footnotesize{}Little difference}} & {\footnotesize{}Slower} & {\footnotesize{}Faster}\tabularnewline
\hline
\end{tabular}

{\footnotesize{}\protect\caption{\label{tab:AIM_CFM_speed}
{\footnotesize{}The speed of the convergence
comparisons between the AIM and the CFM for
calculating the overtone $n=1$ with various $Q
(Q\leq 1)$ and $q$.}} }
\end{table}

Moreover, it is known from
(\ref{eq:AIM_expansion}) that the speed of
convergence of AIM is related to the position of
the expansion point. When $Q$ is small, the
position of the expansion $\xi$ lies in a large
range to permit the convergence of the AIM.
However as the increase of  $Q$, the range of the
allowed expansion point to accommodate
convergence becomes narrower and the value of
$\xi$ needs to be taken smaller. We choose
$\xi=0.45$ when $Q\leq1$ and $\xi=0.43$ when
$Q>1$ in our calculation. See the Table 4 for the
comparison. For $Q>1$, fine tuning of the
expansion point is needed to control the
convergence of the computation by using the AIM.

\begin{table}[h]
\centering
\begin{tabular}{|>{\centering}m{6.5cm}|>{\centering}m{4cm}|>{\centering}m{4cm}|}
\hline
{\footnotesize{}$l=2,Q=1.2,q=2$} & {\footnotesize{}AIM ($\xi=0.45$)} & {\footnotesize{}AIM ($\xi=0.43$)}\tabularnewline
\hline
{\footnotesize{}iterative steps } & {\footnotesize{}80} & {\footnotesize{}80}\tabularnewline
\hline
{\footnotesize{}overtone $n=2$}{\footnotesize \par}

{\footnotesize{}(relative error with respect to 100 steps)} & {\footnotesize{}1.695567-0.631905i}{\footnotesize \par}

{\footnotesize{}($-4.7\times10^{-5},3.3\times10^{-4}$)} & {\footnotesize{}1.695564-0.631910i}{\footnotesize \par}

{\footnotesize{}($-3.9\times10^{-8},5.1\times10^{-7}$)}\tabularnewline
\hline
\end{tabular}

\protect\caption{\label{tab: AIM_expansion}
{\footnotesize{}The dependence of the speed of
the convergence in the AIM on different chosen
positions of the expansion.}}
\end{table}

\section{Numerical results}

In this section we report the frequencies of the
charged scalar perturbations in the stringy black
holes with the change of the  parameters, such as
the charge of the black hole $Q$, the charge of
the perturbing scalar field $q$ and the angular
momentum index $l$. We will analyze the numerical
results to see their effects on the frequencies
of the perturbation.

%%%%%%%%%%%%%%%%%%%%%%%%
%%%%start here%%%%%%%%%%
%%%%%%%%%%%%%%%%%%%%%%%%
\subsection{The fundamental modes}

In this subsection, we study the frequencies of
the fundamental modes of the charged scalar
perturbations. In
Table~\ref{tab:fundamental-modes-L0}, we give the
results for various $Q$ and $q$ when $l=0$. We
use both of the two highly precise methods to
study the frequency domain of the perturbation,
and find that they give consistent results when
$Q \leq 1$. When $Q>1$, for example $Q=1.2$ in
our table, we find that the CFM fails to converge
so that we can only rely on the AIM in the
computation. This challenges the argument that
CFM is the most accurate method in calculating
the frequencies of perturbations
\cite{Cardoso:0905}. In our computations, we fix
the number of iterative steps to be $100$ in the
two methods.

\begin{table}[h]
\centering
\begin{tabular}{|>{\raggedright}m{1.1cm}|>{\centering}p{2.7cm}|>{\centering}p{2.7cm}|>{\centering}p{2.7cm}|>{\centering}p{2.7cm}|}
\hline
\multirow{1}{1.1cm}{{\footnotesize{}$l=0$}} & {\footnotesize{}$Q=0.1$} & {\footnotesize{}$Q=0.5$} & {\footnotesize{}$Q=1$} & {\footnotesize{}$Q=1.2$}\tabularnewline
\hline
{\footnotesize{}$q=0$} & {\footnotesize{}0.11064-0.10493i} & {\footnotesize{}0.11562-0.10599i} & {\footnotesize{}0.13736-0.10961i} & {\footnotesize{}0.15948-0.11120i}\tabularnewline
\hline
{\footnotesize{}$q=0.5$} & {\footnotesize{}0.13026-0.10804i} & {\footnotesize{}0.21716-0.11732i} & {\footnotesize{}0.34938-0.12550i} & {\footnotesize{}0.42072-0.12873i}\tabularnewline
\hline
{\footnotesize{}$q=1$} & {\footnotesize{}0.15019-0.11070i} & {\footnotesize{}0.32485-0.12258i} & {\footnotesize{}0.57549-0.12942i} & {\footnotesize{}0.69534-0.13332i}\tabularnewline
\hline
{\footnotesize{}$q=1.5$} & {\footnotesize{}0.17044-0.11297i} & {\footnotesize{}0.43710-0.12495i} & {\footnotesize{}0.81000-0.12991i} & {\footnotesize{}0.97805-0.13401i}\tabularnewline
\hline
{\footnotesize{}$q=2$} & {\footnotesize{}0.19098-0.11491i} & {\footnotesize{}0.55260-0.12596i} & {\footnotesize{}1.04946-0.12948i} & {\footnotesize{}1.26577-0.13343i}\tabularnewline
\hline
{\footnotesize{}$q=2.5$} & {\footnotesize{}0.21180-0.11658i} & {\footnotesize{}0.67038-0.12634i} & {\footnotesize{}1.29192-0.12885i} & {\footnotesize{}1.55668-0.13252i}\tabularnewline
\hline
{\footnotesize{}$q=3$} & {\footnotesize{}0.23288-0.11800i} & {\footnotesize{}0.78977-0.12643i} & {\footnotesize{}1.53630-0.12827i} & {\footnotesize{}1.84974-0.13159i}\tabularnewline
\hline
{\footnotesize{}$q=4$} & {\footnotesize{}0.27578-0.12025i} & {\footnotesize{}1.03171-0.12630i} & {\footnotesize{}2.02853-0.12735i} & {\footnotesize{}2.43985-0.13001i}\tabularnewline
\hline
{\footnotesize{}$q=5$} & {\footnotesize{}0.31955-0.12188i} & {\footnotesize{}1.27626-0.12607i} & {\footnotesize{}2.52344-0.12673i} & {\footnotesize{}3.03318-0.12886i}\tabularnewline
\hline
{\footnotesize{}$q=6$} & {\footnotesize{}0.36409-0.12306} & {\footnotesize{}1.52236-0.12587i} & {\footnotesize{}3.01986-0.12632i} & {\footnotesize{}3.62839-0.12803i}\tabularnewline
\hline
{\footnotesize{}$q=7$} & {\footnotesize{}0.40928-0.12392i} & {\footnotesize{}1.76944-0.12571i} & {\footnotesize{}3.51721-0.12603i} & {\footnotesize{}4.22478-0.12743i}\tabularnewline
\hline
{\footnotesize{}$q=8$} & {\footnotesize{}0.45505-0.12453i} & {\footnotesize{}2.01718-0.12559i} & {\footnotesize{}4.01517-0.12582i} & {\footnotesize{}4.82197-0.12698i}\tabularnewline
\hline
{\footnotesize{}$q=10$} & {\footnotesize{}0.54797-0.12528i} & {\footnotesize{}2.51392-0.12541i} & {\footnotesize{}5.01226-0.12555i} & {\footnotesize{}6.01789-0.12637i}\tabularnewline
\hline
{\footnotesize{}$q=12$} & {\footnotesize{}0.64236-0.12564i} & {\footnotesize{}3.01169-0.12530i} & {\footnotesize{}6.01027-0.12539i} & {\footnotesize{}7.21530-0.12581i}\tabularnewline
\hline
\end{tabular}

\protect\caption{{\footnotesize{}\label{tab:fundamental-modes-L0}Fundamental
modes of the GHS BH when $l=0$. The CFM and the
AIM give the consistent result when $Q\leq1$.
However, when $Q>1,q=1.2$ in the table for
example, the CFM fails to converge and at this
time we can only rely on the AIM. The iteration
steps in both methods are taken to be 100. The
expansion point in the AIM is taken to be
$\xi=0.43$. For Schwarzschild BH ($Q=0$), the
fundamental mode reads
$\omega=0.11047-0.10487i$.}}
\end{table}

We found that in all cases, the imaginary
frequencies of the perturbations are negative,
which indicate that there are no unstable modes
of the charged scalar perturbation around the GHS
black hole. When the black hole charge
disappears, we recover the Schwarzschild black
hole and the fundamental modes we get in this
limit reproduces the result of the Schwarzschild
black hole under the scalar perturbation. Fixing
the charge of the perturbation field,  we find
that the perturbation presents more oscillations
but decays faster with the increase of the black
hole charge $Q$. Compared with the real part, the
imaginary part of the frequency changes slower
when we vary $Q$ or $q$.

\begin{figure}[!htbp]
\centering
\includegraphics[width=0.45\textwidth]{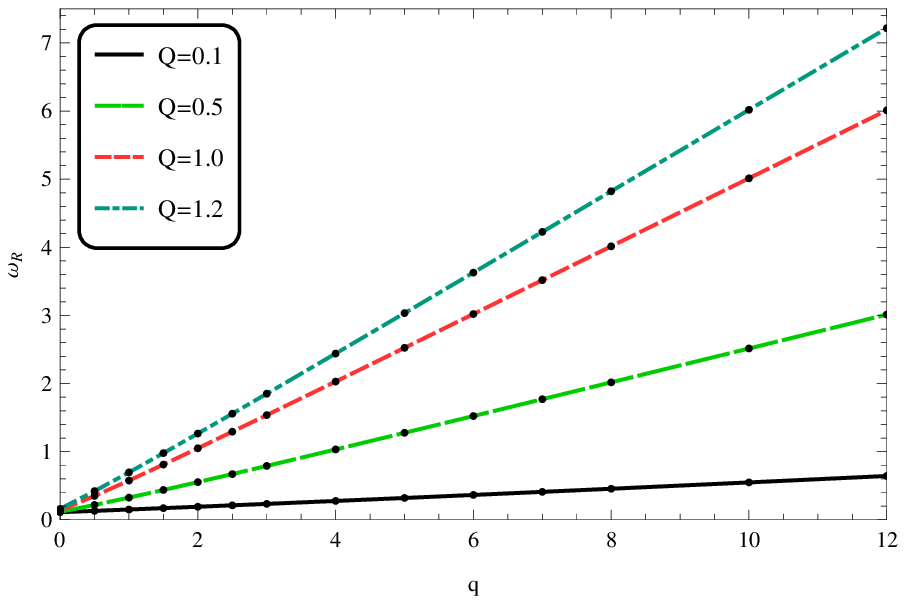}
\includegraphics[width=0.47\textwidth]{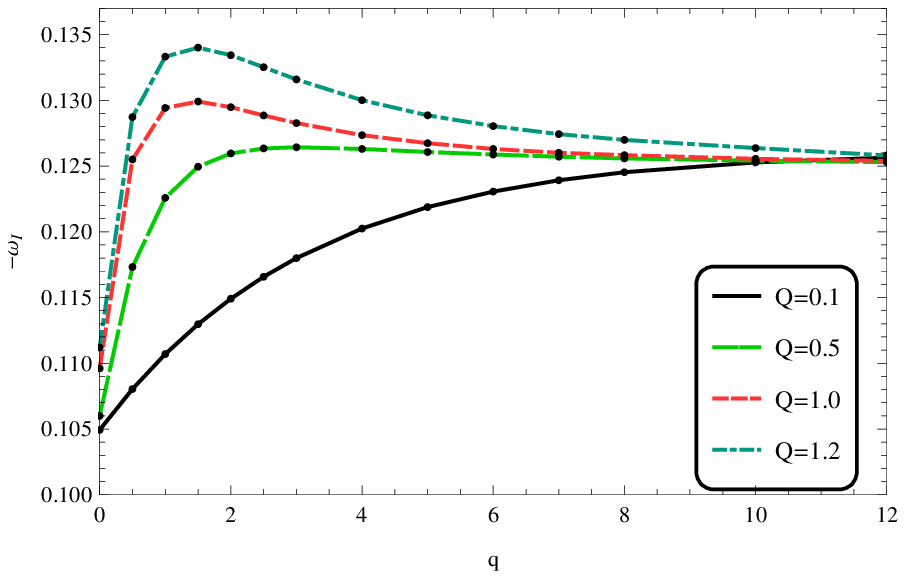}
\caption{\footnotesize{}\label{fig:fundamental-modes-L0} The fundamental
modes of charged scalar perturbation of GHS BH when $l=0$.}
\end{figure}

The objective pictures of the dependence of the
real and imaginary parts of the frequencies on
varying $Q$ and $q$ are shown in
Fig.\ref{fig:fundamental-modes-L0}. From the left
panel, we see that for the fixed $Q$, $\omega_R$
increases linearly as $q$ increases with a slope
$\sim \frac{1}{2} Q$. We fitted the data and
found that $\omega_R \sim 0.079+0.493 Q q$. The
behavior of $\omega_I$ is more complicated. From
the right panel, we observe that when $Q$ is
small, $Q=0.1$ for example, $|\omega_I|$ is a
monotonically increasing function of $q$.
However, when $Q$ becomes large, $|\omega_I|$ is
no longer a  monotonic function of $q$. With the
increase of $q$, $|\omega_I|$ first increases and
then after reaching a maximum value it begins to
decrease. With the increase of $Q$, we find that
the imaginary part of the perturbation becomes
more negative when the perturbation is weakly
charged, which indicates that in this case the
black hole is more stable. When $q$ becomes big
enough, $\omega_I$ converges to a constant no
matter what values of $Q$ one chooses.

\begin{figure}[!htbp]
\centering
\includegraphics[width=0.45\textwidth]{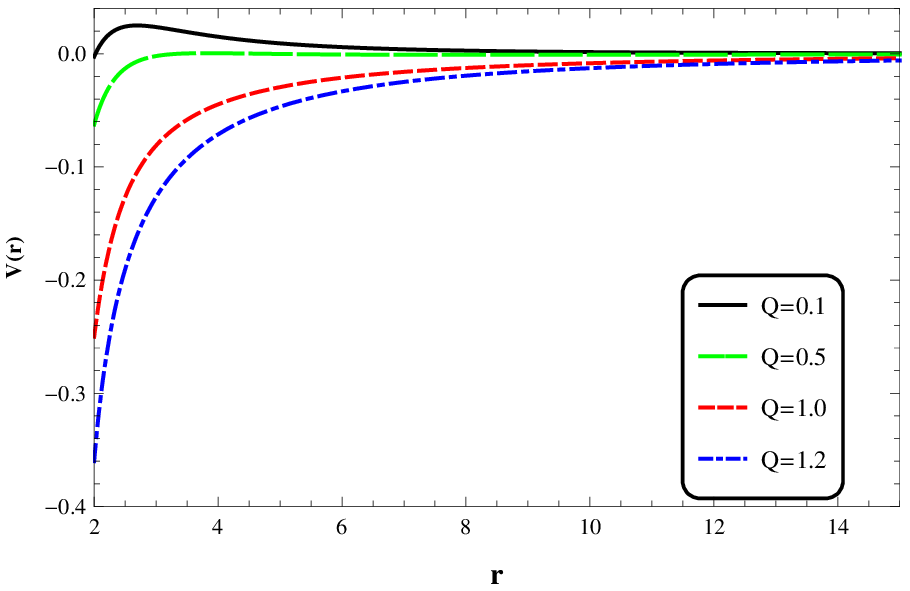}
\includegraphics[width=0.45\textwidth]{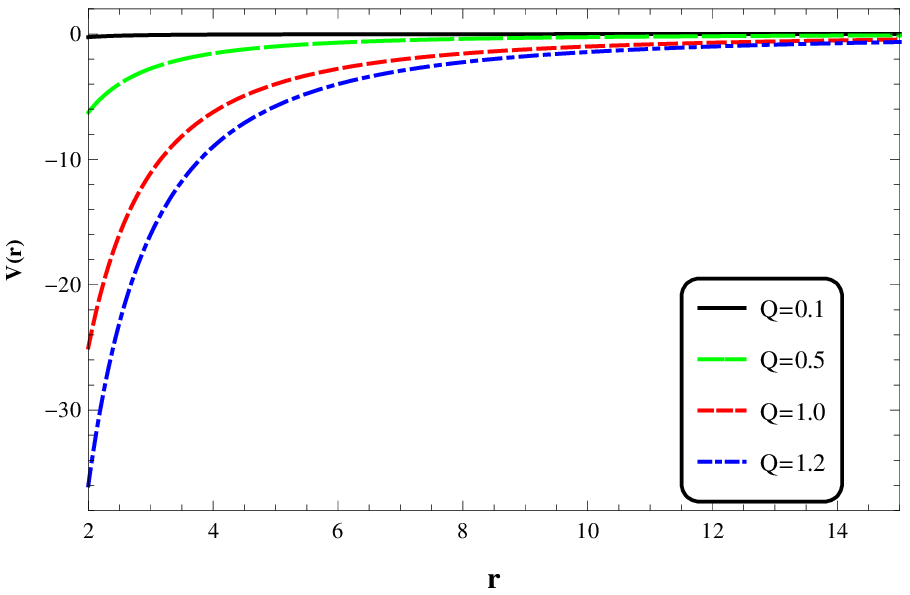}
\caption{\footnotesize{}\label{fig:potential-L0}  The effective potential for different $Q$s when  $l=0$ (left for $q=1$, right for $q=10$).}
\end{figure}

We plot the effective potential in
Fig.\ref{fig:potential-L0} to give some intuitive
understandings of the properties of the
fundamental modes discussed above. We find that
$V_{eff}\rightarrow -\frac{q^{2}Q^{2}}{4M^2}$
when $r\rightarrow 2M$, while $V_{eff}\rightarrow
0$ when $r\rightarrow \infty$. When $q=1$, the
effective potential has a barrier for small $Q$.
The barrier becomes lower and even disappears as
$Q$ increases. This implies that the perturbing
wave can fall into the black hole more easily
when $Q$ becomes larger, which explains the
faster decay of the weakly charged scalar
perturbation with the increase of $Q$ as shown in
the right panel of
Fig.\ref{fig:fundamental-modes-L0}.  On the other
hand, near the horizon, the potential is more
negative for larger $Q$, which tells us that the
perturbation can have more momentum to fall into
the black hole and explains the reason that the
perturbing scalar can have faster oscillation in
the decay when  $Q$ increases. This supports the
observation in the right panel of
Fig.\ref{fig:fundamental-modes-L0} that we have
bigger real part of the perturbation frequency
for bigger $Q$.

When $q$ becomes large enough, $q=10$ for
instance or even bigger, the potential barrier
disappears for all $Q$. The perturbation wave can
be absorbed by the black hole without any
obstacles. This gives the same decay speed of the
perturbation for all values of $Q$ as shown in
the right panel of
Fig.\ref{fig:fundamental-modes-L0}. On the other
hand, at the black hole horizon, the differences
in the potential values there caused by the black
hole charge $Q$ become bigger compared to the
case with weakly charged scalar perturbation.
This explains that the difference in the momentum
for the perturbing wave to fall into the hole is
enlarged when the scalar field is heavily
charged, which explains the big difference in the
real part of the frequency with the change of the
black hole charge $Q$ when the scalar
perturbation is heavily charged as shown in the
left panel of Fig.\ref{fig:fundamental-modes-L0}.

We also perform the same calculations for
different angular indexes $l=1$ and $l=2$,
respectively, to see the effect of the angular
momentum. The results are listed in
Table.\ref{tab:fundamental-modes-L1} and
\ref{tab:overtone_L2}. The overall behaviors of
the fundamental modes for $l=1$ and $l=2$ are
similar to that of $l=0$. Also, we can see that
the imaginary part of the fundamental modes
changes slowly while the real part changes
significantly as we vary $Q$ and $q$.

\begin{table}[!htbp]
\centering
\begin{tabular}{|>{\raggedright}m{1.1cm}|>{\centering}p{2.7cm}|>{\centering}p{2.7cm}|>{\centering}p{2.7cm}|>{\centering}p{2.7cm}|}
\hline
\multirow{1}{1.1cm}{{\footnotesize{}$l=1$}} & {\footnotesize{}$Q=0.1$} & {\footnotesize{}$Q=0.5$} & {\footnotesize{}$Q=1$} & {\footnotesize{}$Q=1.2$}\tabularnewline
\hline
{\footnotesize{}$q=0$} & {\footnotesize{}0.29343-0.09771i} & {\footnotesize{}0.30622-0.09899i} & {\footnotesize{}0.36304-0.10361i} & {\footnotesize{}0.42191-0.10624i}\tabularnewline
\hline
{\footnotesize{}$q=0.5$} & {\footnotesize{}0.31072-0.09940i} & {\footnotesize{}0.39622-0.10623i} & {\footnotesize{}0.55527-0.11470i} & {\footnotesize{}0.66341-0.11785i}\tabularnewline
\hline
{\footnotesize{}$q=1$} & {\footnotesize{}0.32829-0.10100i} & {\footnotesize{}0.49171-0.11158i} & {\footnotesize{}0.76093-0.12112i} & {\footnotesize{}0.91814-0.12457i}\tabularnewline
\hline
{\footnotesize{}$q=1.5$} & {\footnotesize{}0.34613-0.10250i} & {\footnotesize{}0.59173-0.11549i} & {\footnotesize{}0.97615-0.12478i} & {\footnotesize{}1.18197-0.12850i}\tabularnewline
\hline
{\footnotesize{}$q=2$} & {\footnotesize{}0.36422-0.10391i} & {\footnotesize{}0.69548-0.11833i} & {\footnotesize{}1.19842-0.12680i} & {\footnotesize{}1.45242-0.13075i}\tabularnewline
\hline
{\footnotesize{}$q=2.5$} & {\footnotesize{}0.38256-0.10524i} & {\footnotesize{}0.80232-0.12038i} & {\footnotesize{}1.42597-0.12785i} & {\footnotesize{}1.72788-0.13196i}\tabularnewline
\hline
{\footnotesize{}$q=3$} & {\footnotesize{}0.40113-0.10649i} & {\footnotesize{}0.91173-0.12185i} & {\footnotesize{}1.65756-0.12832i} & {\footnotesize{}2.00725-0.13252i}\tabularnewline
\hline
{\footnotesize{}$q=4$} & {\footnotesize{}0.43895-0.10877i} & {\footnotesize{}1.13661-0.12365i} & {\footnotesize{}2.12944-0.12842i} & {\footnotesize{}2.57463-0.13257i}\tabularnewline
\hline
{\footnotesize{}$q=5$} & {\footnotesize{}0.47761-0.11077i} & {\footnotesize{}1.36750-0.12457i} & {\footnotesize{}2.60924-0.12808i} & {\footnotesize{}3.15021-0.13200i}\tabularnewline
\hline
{\footnotesize{}$q=6$} & {\footnotesize{}0.51704-0.11254i} & {\footnotesize{}1.60265-0.12503i} & {\footnotesize{}3.09417-0.12766i} & {\footnotesize{}3.73137-0.13124i}\tabularnewline
\hline
{\footnotesize{}$q=7$} & {\footnotesize{}0.55718-0.11409i} & {\footnotesize{}1.84088-0.12525i} & {\footnotesize{}3.58258-0.12726i} & {\footnotesize{}4.31647-0.13048i}\tabularnewline
\hline
{\footnotesize{}$q=8$} & {\footnotesize{}0.59799-0.11544i} & {\footnotesize{}2.08137-0.12535i} & {\footnotesize{}4.07342-0.12692i} & {\footnotesize{}4.90445-0.12978i}\tabularnewline
\hline
{\footnotesize{}$q=10$} & {\footnotesize{}0.68138-0.11767i} & {\footnotesize{}2.56706-0.12538i} & {\footnotesize{}5.05994-0.12641i} & {\footnotesize{}6.08632-0.12865i}\tabularnewline
\hline
{\footnotesize{}$q=12$} & {\footnotesize{}0.76686-0.11938i} & {\footnotesize{}3.05688-0.12535i} & {\footnotesize{}6.05055-0.12606i} & {\footnotesize{}7.21341-0.12762i}\tabularnewline
\hline
\end{tabular}

\protect\caption{\label{tab:fundamental-modes-L1}{\footnotesize{}Fundamental
modes of the GHS black hole with the angular
index $l=1$. The CFM and the AIM give the same
results when $Q\leq1$. However, when $Q>1,q=1.2$
in the table for example, the CFM fails to
converge and so that we can only rely on the AIM.
The iteration steps in the two methods are taken
to be 100. The expansion point in the AIM is
taken to be $\xi=0.43$. For the limiting case
($Q=0$), we reproduce the result of the
Schwarzschild black hole with the fundamental
mode $\omega=0.29293-0.09766i$. }}
\end{table}

\begin{figure}[!htbp]
\centering
\includegraphics[width=0.45\textwidth]{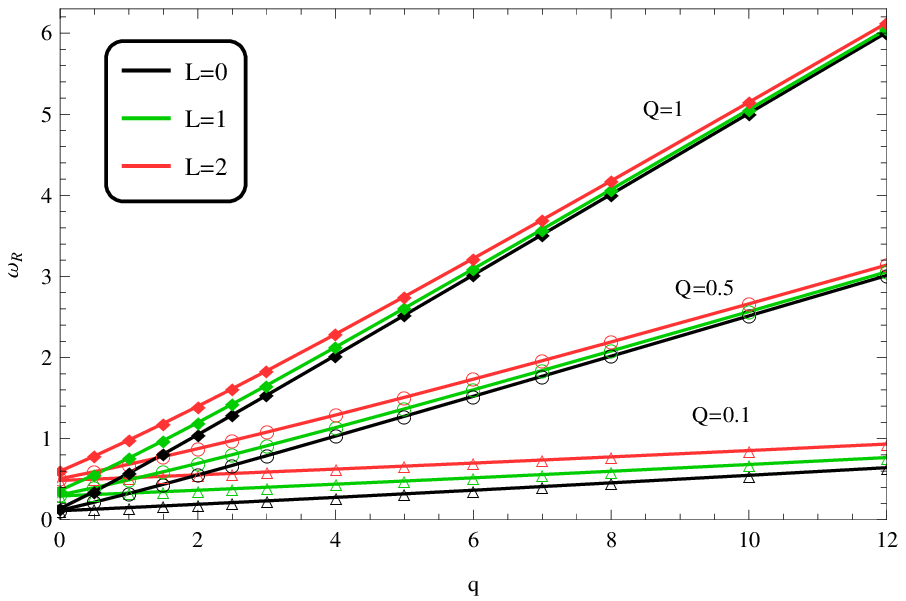}
\includegraphics[width=0.47\textwidth]{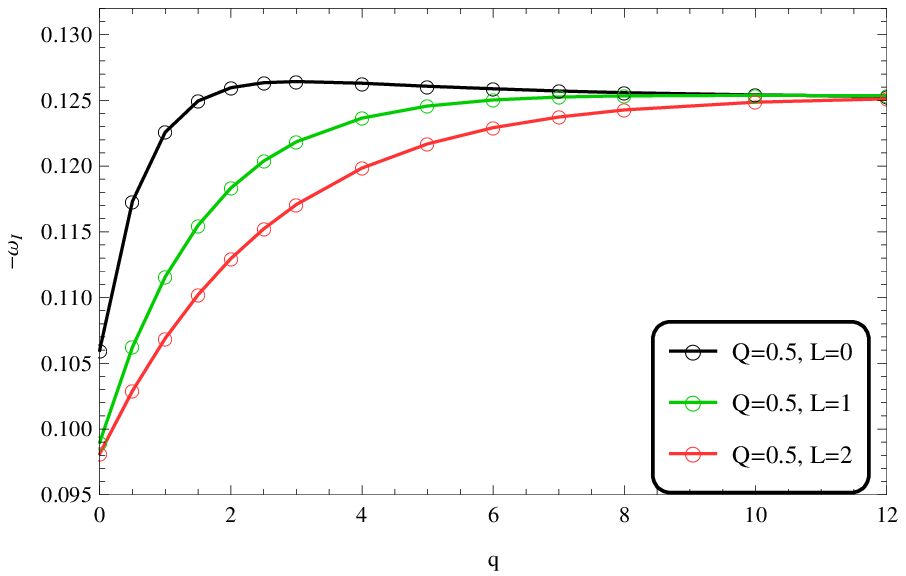}
\caption{\footnotesize{}\label{fig:wRI-fund} The fundamental
modes of GHS BH for different $l$s and $Q$s (The first panel for real parts, the remaining panels for imaginary parts).}
\end{figure}

In Fig.\ref{fig:wRI-fund}, we plot the
fundamental modes for different angular index $l$
($l=0,1,2$). From the figure, we can see that
when $Q$ is fixed, the real part of the frequency
keeps linearly increasing with the increase of
$q$, which is independent of the angular index
$l$. For $l=1$, $\omega_R \sim 0.252+0.481 Q q$.
For $l=2$, $\omega_R \sim 0.455+0.471 Q q$.
Moreover, for fixed $Q$, $\omega_R$s for
different $l$s approach to each other as $q$
becomes very large. This is even clearer when the
black hole charge $Q$ is large.  As $q$
increases, the imaginary part $\omega_I$ flattens
and approaches to constant values regardless of
the chosen $Q$ and $l$.

\begin{figure}[!htbp]
\centering
\includegraphics[width=0.46\textwidth]{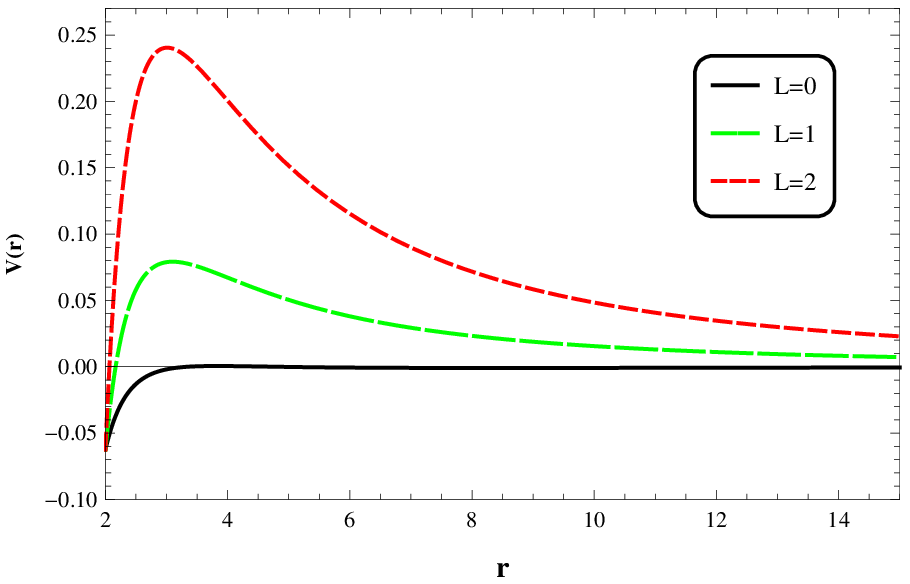}
\includegraphics[width=0.45\textwidth]{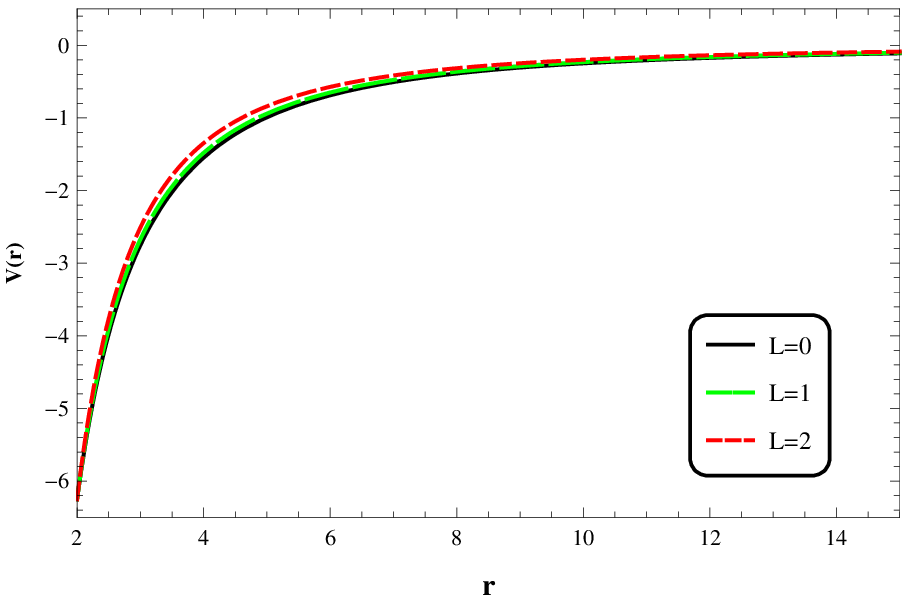}
\caption{\footnotesize{}\label{fig:potential-Ls}  The effective potential for different $l$s when  $Q=0.5$ (left for $q=1$, right for $q=10$).}
\end{figure}

We plot the effective potential for different
$l$s when $Q=0.5$ in Fig.\ref{fig:potential-Ls}
to understand the curves in
Fig.\ref{fig:wRI-fund} more intuitively. When $q$
is small, the potential barrier increases as $l$
increases. This means that the perturbing wave
with higher $l$ is more difficult to be absorbed
into the black hole and decay so that its
imaginary frequency is bigger. At the horizon,
the potential drops faster when $l$ is larger,
thus for higher $l$, the perturbation falls into
the black hole usually with bigger momentum so
that it can  oscillate faster. At large $q$, the
potential barriers disappear for all values of
$l$. Furthermore, they basically coincide with
each other for large $q$ ($q=10$ for example in
Fig.\ref{fig:potential-Ls}). So both the real
parts and the imaginary parts of the perturbation
modes approach to each other at large $q$, as
shown in Fig.\ref{fig:wRI-fund}.

\subsection{Overtones}

Now we turn to present the results of the
overtones of the charged scalar perturbation
around the GHS black hole. We only list the
fundamental modes and the first two overtones
when $l=2$ in Table.\ref{tab:overtone_L2} for
concision.

\begin{table}[H]
\centering
\begin{tabular}{|>{\raggedright}m{1.1cm}|>{\centering}m{1.4cm}|>{\centering}m{2.7cm}|>{\centering}m{2.7cm}|>{\centering}m{2.7cm}|>{\centering}m{2.7cm}|}
\hline
\multirow{1}{1.1cm}{{\footnotesize{}$l=2$}} & {\footnotesize{}overtones} & {\footnotesize{}$Q=0.1$} & {\footnotesize{}$Q=0.5$} & {\footnotesize{}$Q=1$} & {\footnotesize{}$Q=1.2$}\tabularnewline
\hline
{\footnotesize{}$q=0$} & {\footnotesize{}$n=0$}{\footnotesize \par}

{\footnotesize{}$n=1$}{\footnotesize \par}

{\footnotesize{}$n=2$} & {\footnotesize{}0.48445-0.09681i}{\footnotesize \par}

{\footnotesize{}0.46470-0.29575i}{\footnotesize \par}

{\footnotesize{}0.43145-0.50876i} & {\footnotesize{}0.50541-0.09812i}{\footnotesize \par}

{\footnotesize{}0.48660-0.29942i}{\footnotesize \par}

{\footnotesize{}0.45489-0.51401i} & {\footnotesize{}0.59878-0.10291i}{\footnotesize \par}

{\footnotesize{}0.58380-0.31271i}{\footnotesize \par}

{\footnotesize{}0.55839-0.53288i} & {\footnotesize{}0.69577-0.10572i}{\footnotesize \par}

{\footnotesize{}0.68476-0.32025i}{\footnotesize \par}

{\footnotesize{}0.66619-0.54286i}\tabularnewline
\hline
{\footnotesize{}$q=2$} & {\footnotesize{}$n=0$}{\footnotesize \par}

{\footnotesize{}$n=1$}{\footnotesize \par}

{\footnotesize{}$n=2$} & {\footnotesize{}0.55330-0.10086i}{\footnotesize \par}

{\footnotesize{}0.53702-0.30696i}{\footnotesize \par}

{\footnotesize{}0.50965-0.52437i} & {\footnotesize{}0.87667-0.11296i}{\footnotesize \par}

{\footnotesize{}0.86989-0.34051i}{\footnotesize \par}

{\footnotesize{}0.85816-0.57223i} & {\footnotesize{}1.39915-0.12269i}{\footnotesize \par}

{\footnotesize{}1.39785-0.36838i}{\footnotesize \par}

{\footnotesize{}1.39555-0.61487i} & {\footnotesize{}1.69259-0.12634i}{\footnotesize \par}

{\footnotesize{}1.69356-0.37907i}{\footnotesize \par}

{\footnotesize{}1.69556-0.63191i}\tabularnewline
\hline
{\footnotesize{}$q=4$} & {\footnotesize{}$n=0$}{\footnotesize \par}

{\footnotesize{}$n=1$}{\footnotesize \par}

{\footnotesize{}$n=2$} & {\footnotesize{}0.62484-0.10437i}{\footnotesize \par}

{\footnotesize{}0.61152-0.31664i}{\footnotesize \par}

{\footnotesize{}0.58903-0.53793i} & {\footnotesize{}1.29143-0.11985i}{\footnotesize \par}

{\footnotesize{}1.28926-0.35996i}{\footnotesize \par}

{\footnotesize{}1.28531-0.60114i} & {\footnotesize{}2.28913-0.12753i}{\footnotesize \par}

{\footnotesize{}2.28956-0.38258i}{\footnotesize \par}

{\footnotesize{}2.29043-0.63758i} & {\footnotesize{}2.77507-0.13167i}{\footnotesize \par}

{\footnotesize{}2.77690-0.39496i}{\footnotesize \par}

{\footnotesize{}2.78057-0.65810i}\tabularnewline
\hline
{\footnotesize{}$q=6$} & {\footnotesize{}$n=0$}{\footnotesize \par}

{\footnotesize{}$n=1$}{\footnotesize \par}

{\footnotesize{}$n=2$} & {\footnotesize{}0.69884-0.10739i}{\footnotesize \par}

{\footnotesize{}0.68798-0.32498i}{\footnotesize \par}

{\footnotesize{}0.66956-0.54970i} & {\footnotesize{}1.73335-0.12292i}{\footnotesize \par}

{\footnotesize{}1.73272-0.36885i}{\footnotesize \par}

{\footnotesize{}1.73153-0.61501i} & {\footnotesize{}3.22242-0.12824i}{\footnotesize \par}

{\footnotesize{}3.22267-0.38472i}{\footnotesize \par}

{\footnotesize{}3.22316-0.64124i} & {\footnotesize{}3.89915-0.13246i}{\footnotesize \par}

{\footnotesize{}-}{\footnotesize \par}

{\footnotesize{}-}\tabularnewline
\hline
{\footnotesize{}$q=8$} & {\footnotesize{}$n=0$}{\footnotesize \par}

{\footnotesize{}$n=1$}{\footnotesize \par}

{\footnotesize{}$n=2$} & {\footnotesize{}0.77505-0.10998i}{\footnotesize \par}

{\footnotesize{}0.76626-0.33215i}{\footnotesize \par}

{\footnotesize{}0.75120-0.55992i} & {\footnotesize{}2.19246-0.12427i}{\footnotesize \par}

{\footnotesize{}2.19230-0.37283i}{\footnotesize \par}

{\footnotesize{}2.19200-0.62143i} & {\footnotesize{}4.17893-0.12791i}{\footnotesize \par}

{\footnotesize{}4.17894-0.38374i}{\footnotesize \par}

{\footnotesize{}4.17898-0.63962i} & \tabularnewline
\hline
{\footnotesize{}$q=10$} & {\footnotesize{}$n=0$}{\footnotesize \par}

{\footnotesize{}$n=1$}{\footnotesize \par}

{\footnotesize{}$n=2$} & {\footnotesize{}0.85328-0.11220i}{\footnotesize \par}

{\footnotesize{}0.84618-0.33831i}{\footnotesize \par}

{\footnotesize{}0.83392-0.56879i} & {\footnotesize{}2.66269-0.12486i}{\footnotesize \par}

{\footnotesize{}2.66267-0.37459i}{\footnotesize \par}

{\footnotesize{}2.66262-0.62432i} & {\footnotesize{}5.14884-0.12741i}{\footnotesize \par}

{\footnotesize{}5.14876-0.38224i}{\footnotesize \par}

{\footnotesize{}5.14859-0.63710i} & \tabularnewline
\hline
{\footnotesize{}$q=12$} & {\footnotesize{}$n=0$}{\footnotesize \par}

{\footnotesize{}$n=1$}{\footnotesize \par}

{\footnotesize{}$n=2$} & {\footnotesize{}0.93333-0.11409i}{\footnotesize \par}

{\footnotesize{}0.92763-0.34359i}{\footnotesize \par}

{\footnotesize{}0.91766-0.57647i} & {\footnotesize{}3.14034-0.12511i}{\footnotesize \par}

{\footnotesize{}3.14034-0.37534i}{\footnotesize \par}

{\footnotesize{}3.14036-0.62557i} & {\footnotesize{}6.12702-0.12696i}{\footnotesize \par}

{\footnotesize{}6.12691-0.38089i}{\footnotesize \par}

{\footnotesize{}6.12669-0.63482i} & \tabularnewline
\hline
\end{tabular}

\protect\caption{\label{tab:overtone_L2}{\footnotesize{}The
frequencies of the charged scalar perturbations
around the GHS black hole when $l=2$. The CFM and
the AIM give the same result when $Q\leq1$.
However, the CFM fails when $Q>1$. The
frequencies for $Q=1.2$ are calculated by the
AIM. We adjust the expansion point to keep high
precision and efficiency. The iteration step is
100. When $Q = 1.2$ and $q
> 6$, the precision of the results is lower than
$10^{-5}$ and the results are not shown in this
table. For the Schwarzschild black hole ($Q=0$),
the fundamental mode reads
$\omega=0.48364-0.09675i$.}}
\end{table}

It can be seen from Table.\ref{tab:overtone_L2}
that when $Q,q$ and $l$ are fixed, the real parts
of the overtones are nearly invariant while the
imaginary parts change more significantly. To see
more clearly, we plot the overtones for $l=1$ and
$l=2$ in Fig. 5. We can see that the real part
$\omega_R$ is nearly independent of $n$, while
the imaginary part $\omega_I$ depends on $n$
linearly. For example, when $l=1$, $\omega_I$ can
be approximated well by the form $\omega_I \sim
0.1262+0.2562n$ when $q=6$ and $\omega_I \sim
0.1261+0.2520n$ when $q=12$. As $q$ increases,
the slope tends to $0.252$. Meanwhile, the modes
for different $l$ also become closer to each
other as $q$ increases. This can be understood
since the effective potential for different $l$
approaches to each other when $q$ is large.

\begin{figure}[H]
\begin{longtable}{l||l}
\multicolumn{2}{l}{\includegraphics[scale=0.8]{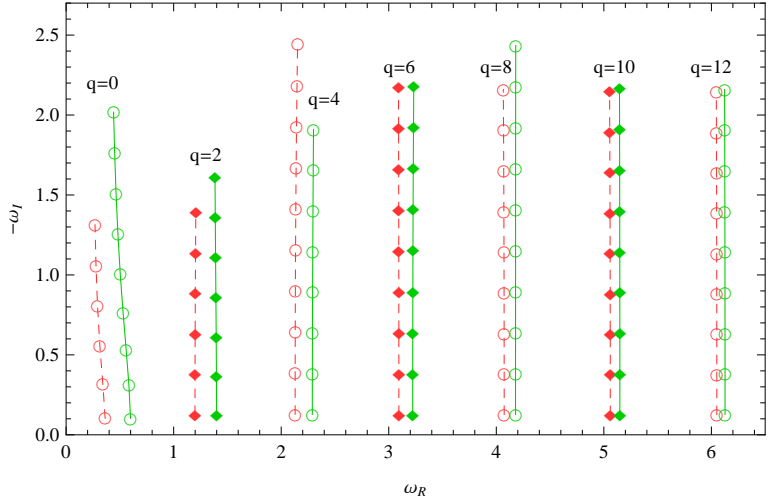}}\tabularnewline
\end{longtable}

\protect\caption{\label{fig:overtones-GHS}{\footnotesize{}Overtones
of perturbations around GHS black hole when
$Q=1$. The dashed left red line corresponds to
the overtones when $l=1$ and the right green line
corresponds to the overtones when $l=2$. }}
\end{figure}

\subsection{The superradiance and its stability}

In this subsection we will show that the
superradiance with the amplification of the
incident wave can not appear for the GHS black
hole. We consider the classical scattering
problem for a charged scalar field in the GHS
black hole background.

The asymptotic behavior of (\ref{eq:radial_eq})
when $\omega^{2}>\mu^{2}$ can be derived
straightforwardly,
\begin{eqnarray}
\Psi & \sim & \begin{cases}
Te^{-i\left(\omega-\frac{qQ}{2M}\right)r_{*}}, & r_{*}\rightarrow-\infty(r\rightarrow2M),\\
e^{-i\sqrt{\omega^{2}-\mu^{2}}r_{*}}+Be^{i\sqrt{\omega^{2}-\mu^{2}}r_{*}},
& r_{*}\rightarrow\infty(r\rightarrow\infty).
\end{cases}\label{eq:boundary}
\end{eqnarray}
This boundary condition corresponds to an
incident wave of unit amplitude from the infinity
and a reflected wave of amplitude $B$ back to the
infinity and a transmitted wave of amplitude $T$
towards the event horizon. Since the effective
potential is real, $\Psi^{*}$ is also a solution
of the radial equation Eq.~(\ref{eq:radial_eq})
and independent of $\Psi$. Thus, the Wronskian $W
\equiv
\Psi\frac{d}{dr_{*}}\Psi^{*}-\Psi^{*}\frac{d}{dr_{*}}\Psi$
is independent of $r_\ast$. Calculating the
Wronskian at the black hole horizon and at the
infinity respectively, and equaling the two
values, we get
\begin{eqnarray}
1-|B|^{2} & = &
\frac{\omega-\frac{qQ}{2M}}{\sqrt{\omega^{2}-\mu^{2}}}|T|^{2}.
\end{eqnarray}
We see that if $\mu<\omega<\frac{qQ}{2M}$, the
amplitude of the reflected wave is larger than
the one of the incident wave $|B|^{2}>1$. This
phenomenon is known as superradiance. Thus, we
get the superradiance condition
\begin{eqnarray}\label{superadiance_condition}
\mu<\omega<\frac{qQ}{2M}.
\end{eqnarray}
This condition was also derived
in~\cite{Li:2013ms,Shiraishi:1992st}.

Multiplying the complex conjugated field
$\Psi^{*}$ on both sides of (\ref{eq:radial_eq})
and doing partial integration, we get
\begin{equation}
\Psi^{*}(r_{*})\Psi'(r_{*})|_{-\infty}^{\infty}+\int_{-\infty}^{\infty}\left(\omega-q\Phi\right)^{2}|\Psi(r_{*})|^{2}dr_{*}=\int_{-\infty}^{\infty}\left(V|\Psi(r_{*})|^{2}+|\Psi'(r_{*})|^{2}\right)dr_{*},
\end{equation}
 in which $\Phi=\frac{Q}{r}$. Note that the right hand side of the above equation is real and positive
since $V$ is positive. Taking imaginary part of
both sides, we get
\begin{equation}
\left(a^{2}+b^{2}\right)^{\frac{1}{4}}\cos\left(\frac{1}{2}\arctan\frac{b}{a}\right)+\omega_{R}-q\Phi_{h}+\int_{-\infty}^{\infty}2\omega_{I}\left(\omega_{R}-q\Phi\right)|\Psi(r_{*})|^{2}dr_{*}=0.
\end{equation}
Here we have taken notations
$\omega=\omega_{R}+i\omega_{I}$,
$a=\omega_{R}^{2}-\omega_{I}^{2}-\mu^{2}$,
$b=2\omega_{R}\omega_{I}$ and
$\Phi_h=\frac{qQ}{2M}$. Since
$\arctan\frac{b}{a}\in(-\frac{\pi}{2},\frac{\pi}{2})$
and $\Phi$ is a monotonic decreasing function, we
get $\omega_{I}<0$ if $\omega_{R}>q\Phi_{h}$.
Thus, if the instability with $\omega_{I}>0$
occurs, we must have $\omega_{R}<q\Phi_{h}$. This
tells us that the instability can only take place
when
\begin{equation}
\mu<\omega_{R}<q\Phi_{h},
\end{equation}
which is the superradiance condition
Eq.~(\ref{superadiance_condition}). Thus if the
GHS black hole experiences instability, this
instability  must be superradiant instability.

But from the above computations, we always find
that the perturbation modes with $\omega_R$ are
beyond the superradiant condition
Eq.~(\ref{superadiance_condition}), no matter
what parameter ranges we choose. According to the
analytical argument, these modes should be stable
with $\omega_I<0$, which are consistent with our
numerical results above. On the other hand, from
the effective potential, we do not see the
potential well outside the black hole to
accumulate the energy. The necessary condition
for the existence of the superradiant instability
does not hold. Thus the superradiant instability
cannot exist in the GHS background.

\section{Summary and discussion}

In this paper, we have discussed the stability of
the GHS black hole under charged scalar
perturbations. For the charged scalar
perturbations, we have found that not all
available numerical methods can efficiently
compute the accurate frequencies of the
perturbations. This is different from that of the
neutral scalar perturbations. We have discovered
that two numerical methods, the CFM and the AIM,
can still keep high accuracy and efficiency in
the computations of the frequency of the charged
scalar perturbations. We have experienced that
the speed of convergence of the AIM depends on
the position of the expansion point which needs
to be chosen suitably. Besides, we have observed
that the speed of convergence of CFM is close to
that of the AIM when the black hole charge $Q$ is
small. However, when $Q$ becomes large but still
smaller than the mass of the black hole $M$, the
AIM has faster convergence than the CFM when the
scalar field is weakly charged; but the result is
opposite when the charge of the scalar field is
big.  When the black hole charge $Q$ exceeds the
black hole mass $M$, the CFM is found invalid to
give reliable results, while the AIM can still
work. However, the convergence of the AIM becomes
bad as $Q$ approaches to the extremal value
$\sqrt{2}$. The results have been summarized in
Table 3. The comparisons of the efficiency and
accuracy among different numerical methods are
important, because the accuracy and convergence
of the numerical computations are key
requirements to grasp the properties in the
perturbations around black holes.

The influences on the frequencies of the
perturbations brought by parameters describing
the background and the perturbation field have
been illustrated.  The intuitive reasons behind
these phenomena have also been discussed. We have
observed that the GHS black hole spacetime, which
can reduce to the Schwarzschild black hole when
the black hole charge goes to zero, is stable
against the charged scalar perturbation. This
result is different from what have been disclosed
in the AdS black holes and the dS black holes
with vanishing angular momentum, where the
backgrounds experience instability under the
charged scalar perturbations, while keep stable
against neutral scalar perturbations.

In \cite{Zhu:2014} it was concluded that the new
instability in the dS black hole against charged
scalar perturbations with vanishing angular
momentum is caused by the superradiance. This was
further confirmed in \cite{Konoplya:2014lha}. For
the GHS black hole spacetime, we have shown that
the superradiance does not happen. Recently the
superradiant instability of the GHS background
was discussed in \cite{Li:2014,Li:2014fna}. But
they put the black hole in the artificial cavity.
Although in this way they claimed that they can
devise a black hole bomb, the mechanism with the
artificial mirror is not convincing. One needs to
introduce a natural wall, for example the massive
fields \cite{hod, Zhang:2013haa}, to trigger the
superradiant instability. It was proved in
\cite{Li:2013ms} that the reflecting mirror made
by the mass term of the incident field cannot
trigger the superradiant instability and create
the GHS black hole bomb. Our result has further
argued that the GHS black hole background is
always stable against the charged scalar
perturbation. Considering that the incident
charged scalar wave cannot be amplified due to
the superradiance around the GHS hole and the
property of the effective potential, we find that
the superradiant instability cannot happen in the
GHS black hole background.

\section*{Acknowledgements}

We thank E. Abdalla, Y.Q. Liu, Z.Y. Zhu and D.C.
Zou for helpful discussions. This work was
supported by NNSF of China.

\end{document}